\documentclass[journal]{IEEEtran}
\usepackage{cite}
\ifCLASSINFOpdf
\usepackage[pdftex]{graphicx}
\DeclareGraphicsExtensions{.pdf,.jpeg,.png}
\else
\usepackage[dvips]{graphicx}
\DeclareGraphicsExtensions{.eps}
\fi
\usepackage{epsfig,epsf,epstopdf,graphicx}
\usepackage[cmex10]{amsmath}
\usepackage{amssymb}
\usepackage{amsthm }
\usepackage{nicefrac}
\usepackage{hyperref}
\usepackage{xcolor}
\hypersetup{colorlinks=true}
\usepackage{tabularx}

\theoremstyle{theorem}

\theoremstyle{definition}

\theoremstyle{plain}

\theoremstyle{plain}

\newcommand{\onevec}{{\bf{1}}}
\newcommand{\xb}{{\textbf{x}}}
\newcommand{\yb}{{\textbf{y}}}

\newcommand{\Bb}{{\textbf{B}}}

\newcommand{\Db}{{\textbf{D}}}

\newcommand{\eb}{{\textbf{e}}}

\newcommand{\Ab}{{\textbf{A}}}

\newcommand{\Ib}{{\textbf{I}}}

\newcommand{\rb}{{\textbf{r}}}

\newcommand{\Pb}{{\textbf{P}}}

\newcommand{\thetab}{{\mbox{\boldmath $\theta$}}}

\newcommand{\Lb}{\mathbf{L}}

\newcommand{\Wb}{\mathbf{W}}
\usepackage{algorithm}
\usepackage{multirow}
\usepackage{algcompatible}
\usepackage[bottom]{footmisc}
\PassOptionsToPackage{dvipsnames}{xcolor}
\usepackage{tikz}
\usetikzlibrary{calc}

\ifCLASSOPTIONcompsoc
  \usepackage[caption=false,font=normalsize,labelfont=sf,textfont=sf]{subfig}
\else
  \usepackage[caption=false,font=footnotesize]{subfig}
\fi
\usepackage{fixltx2e}
\usepackage{afterpage}
\usepackage{float}
\usepackage{stfloats}

\makeatletter

\begin{document}

\title{Statistical Graph Signal Recovery Using Variational Bayes}

\author{Razieh~Torkamani,
Hadi~Zayyani,%~\IEEEmembership{Member,~IEEE,}
%Mario~A.~T.~Figueiredo,~\IEEEmembership{Fellow,~IEEE,}
%        and~Farrokh~Marvasti,~\IEEEmembership{Life Senior Member,~IEEE}% <-this % stops a space
\thanks{This work was supported by the Iran National Science Foundation (INSF) (grant number
98003594).}% <-this %
\thanks{R.~Torkamani is with the Department
of Electrical Engineering, K.N. Toosi University of Technology, Tehran, Iran (e-mail: rtorkamani@mail.kntu.ac.ir).}
\thanks{H.~Zayyani is with the Department
of Electrical and Computer Engineering, Qom University of Technology (QUT), Qom, Iran (e-mail: zayyani@qut.ac.ir).}% <-this %
\vspace{-0.5cm}}


\maketitle
\thispagestyle{plain}
\pagestyle{plain}

\begin{abstract}
This paper investigates the problem of graph signal recovery (GSR) when the topology of the graph is not known in advance. In this paper, the elements of the weighted adjacency matrix is statistically related to normal distribution and the graph signal is assumed to be Gaussian Markov Random Field (GMRF). Then, the problem of GSR is solved by a Variational Bayes (VB) algorithm in a Bayesian manner by computing the posteriors in a closed form. The posteriors of the elements of weighted adjacency matrix are proved to have a new distribution which we call it generalized compound confluent hypergeometric (GCCH) distribution. Moreover, the variance of the noise is estimated by calculating its posterior via VB. The simulation results on synthetic and real-world data shows the superiority of the proposed Bayesian algorithm over some state-of-the-art algorithms in recovering the graph signal.

\end{abstract}

\begin{IEEEkeywords}
Graph signal recovery, Laplacian matrix, Sampling, Variational Bayes, Generalized CCH distribution.
\end{IEEEkeywords}

% \fi
\IEEEpeerreviewmaketitle

\section{Introduction}
\label{sec:Intro}
\IEEEPARstart{I}{n} recent years, the Graph Signal Processing (GSP) is a new paradigm to model and analyze the modern data which is defined on a graph \cite{Sand13}, \cite{Shum13}, \cite{GSP18}. GSP has found many applications in sensor networks \cite{GSP18}, biological networks \cite{GSP18}, image processing and machine learning  \cite{GSP18}, \cite{Sadr18}, and power systems \cite{Dine19}. Many problems are arisen in GSP which Graph Signal Recovery (GSR) \cite{Chen15}-\cite{Brug20} , graph signal denoising \cite{Tay20}, and graph learning \cite{Dong16}-\cite{Rame19} to name a few. One of the basic problems in GSP is GSR wherein the graph signal should be reconstructed from its noisy sampled measurements. In \cite{Chen15}, a GSR algorithm based on variation minimization is suggested for recovery of smooth graph signals. Moreover, a distributed least square reconstruction algorithm is proposed to recover the unknown bandlimited graph signal iteratively \cite{Wang15}. In other work \cite{Wang15may}, the concept of local set is introduced and two local-set-based iterative algorithms are devised to recover the bandlimited graph signal. In addition, in \cite{Wang16}, the bandlimited graph signal is reconstructed perfectly from the local measurements. Also, adaptive algorithms are proposed to recover the graph signal in an adaptive manner \cite{Lore16}, \cite{Lore17}, \cite{Lore18}. Besides, a batch reconstruction method of time-varying graph signal is proposed by exploiting the smoothness of the temporal difference \cite{Qiu17}. %Moreover, a first-order primal-dual algorithm is proposed to solve the GSR problem using the total variation minimization \cite{Berg17}. Finally, \cite{Rome17} formulates the graph signal reconstruction as a regression task on reproducing kernel Hilbert spaces of graph signals.

In the aforementioned papers \cite{Chen15}-\cite{Qiu17}, it is assumed that the topology of graph is known in advance. But, in many applications, the underlying graph of the signal is not ready beforehand of processing the signal. Therefore, a graph topology learning process is needed. Hence, some topology learning and graph signal recovery algorithms are suggested to simultaneously recover the graph signal and infer the graph topology \cite{Dong16}-\cite{Rame19}. In \cite{Dong16}, a factor analysis model is adopted for the graph signals and a Gaussian probabilistic prior is used on the latent variables. Then, a learning algorithm is proposed to both recover the smooth graph signal and learn the Laplacian matrix. In a recent work \cite{Ioan19}, two structural model are suggested for graph signal and the topology of the graph is learned via introducing a regularized least-squares optimization problem. Also, in other recent paper \cite{Rame19}, a Bayesian approach is suggested for graph topology learning and signal recovery. In this work, a factor analysis model is used for the graph signal model and a minimum mean square error estimator and an Expectation-Maximization (EM) algorithm are utilized. In a more recent work \cite{Rey19}, a more complex model is used for graph signal which is diffused sparse graph signals. In this paper, a sampling and reconstruction method is proposed for recovery of these graph signals. In addition, in \cite{Wang18}, a robust signal reconstruction algorithm is suggested which is proved to be more robust than conventional least square recovery. Finally, recently, an iterative algorithm is proposed to interpolate graph signals from only a partial set of samples \cite{Brug20}.

In this letter, a GMRF model is used for graph signals and the unknown weighted adjacency matrix elements are related to a statistical Gaussian distribution. Then, the problem of GSR is solved in a Bayesian manner and using a VB framework. As far as we know, the statistical assumption on weighted adjacency matrix elements and VB approach is not utilized beforehand in the literature of GSP. So, the contributions of this paper are 1) The statistical assumption on Laplacian matrix elements and statistical graph signal recovery and Laplacian matrix estimation based on that. 2) Using a Bayesian method via variational Bayes algorithm. 3) Posterior calculations in a closed form formula and defining a new distribution called generalized CCH (GCCH). 4) Noise variance estimation in a Bayesian manner. Also, simulation experiments on both synthetic graph signals and real-world temperature data on a sensor network show the superiority of the proposed VB signal recovery over some state-of-the-art competing algorithms.

%In addition, the posteriors of the Laplacian matrix elements are calculated in a closed form and is proved to have a new distribution which we call it generalized CCH (GCCH) distribution. Moreover, the posterior of the noise variance is calculated and the noise variance is also estimated in a Bayesian manner. Simulation experiments on both synthetic graph signals and real-world temperature data on a sensor network show the superiority of the proposed VB signal recovery over some state-of-the-art competing algorithms.

\section{Problem Formulation}
\label{sec:ProblemForm}

The problem which is investigated in this paper is to recover a graph signal given a subset of samples of the graph. Suppose there is a graph $G=(\cal V,\cal E)$ with $N$ vertices collected in $\cal V$$=\{v_1,v_2,...,v_N\}$ and $\cal E$ is the set of edges. The degree matrix of the graph is defined as $\Db=\mathrm{diag}(\Wb.\onevec_N)$ where $\Wb$ is the weighted adjacency matrix of the graph. The Laplacian matrix over the graph is $\Lb=\Db-\Wb$. Assume that $M_k$ vertices are selected randomly at time $k$ $(1\le k\le K)$, and the graph signal is sampled on that vertices. Then, $k$'th observation can be written as $\yb_k=\Ab_k\xb_k+\eb_k$ where $\xb_k\in\mathbf{R}^{N\times 1}$ is the graph signal at time $k$, $\eb_k\in\mathbf{R}^{M_k\times 1}$ is the measurement noise vector, and $\Ab_k\in\mathbf{R}^{M_k\times N}$ is a matrix in which each row has only a ``one'' element related to the sampled node, and the other components are zero-valued. If we arrange the observations $\{\yb_k\}^{K}_{k=1}$ in one column, we can constitute an $\cal{M}\mathrm{\times 1}$ column vector $\yb$, i.e. $\yb=[\yb^T_1,...,\yb^T_K]^T$, where $\cal{M}\mathrm{=\sum_{k=1}^K M_k}$. Also, the sampling matrices $\Ab_k$, unknown graph signals $\xb_k$, and the measurement noises $\eb_k, 1\le k\le K$ can be gathered in the following model
%matrix $\Psi=\mathrm{diag}(\Ab_i)$, column vectors $\xb=[\xb^T_1,...,\xb^T_K]^T$ and $\eb=[\eb_1^T,...,\eb^T_K]^T$, respectively. %Therefore, the corresponding model is
\begin{equation}
\label{eq: model}
\yb=\Psi\xb+\eb
\end{equation}
where $\xb=[\xb^T_1,...,\xb^T_K]^T\in\mathbf{R}^{NK\times 1}$, $\eb=[\eb_1^T,...,\eb^T_K]^T\in\mathbf{R}^{\cal{M}\mathrm{\times 1}}$, and $\Psi=\mathrm{diag}(\Ab_1,\Ab_2,...,\Ab_K)\in\mathbf{R}^{\cal{M}\mathrm{\times NK}}$.
The problem is to recover the collected graph signal $\xb$ from the sampled set of observations $\yb$. In this paper, the Laplacian matrix of the graph is assumed to be unknown and we learn it in the algorithm in a statistical manner.

\section{The proposed graph signal recovery algorithm}

\subsection{Statistical models}
\label{sec: stat}
The main idea of this letter is to statistically model the Laplacian matrix and the graph signal, and then, estimate the posterior distributions via VB inference. Similar to the literature of statistical GSP \cite{Zhang15}, \cite{Segar18}, we use a Gaussian markov random filed (GMRF) for the collected graph signal $\xb$. So, we have:
\begin{equation}
p(\xb)=(2\pi)^{-\frac{NK}{2}}|\Bb|^{\frac{1}{2}}\exp(-\frac{1}{2}\xb^T\Bb\xb)
\end{equation}
where $\Bb$ is the precision matrix of the graph signal, and can be written as $\Bb=\Ib_{K}\otimes (\Lb+\epsilon \Ib)$ \cite{Sakiyama19}, in which the parameter $\epsilon$ is used to avoid the singularity of the precision matrix and the operator $\otimes$ is Kronecker product. If we nominate $\ell_{ij}$ and $w_{ij}$ as the $(i,j)$'th element of the Laplacian matrix $\Lb$ and the weighted adjacency matrix, respectively, then we have $\ell_{ij}=-w_{ij}$ for all $i\neq j$. In order to assign a prior distribution for the non-diagonal elements of the Laplacian matrix, we use this direct relationship between the elements of the Laplacian and the weighted adjacency matrices, and consider a statistical model for $w_{ij}$, instead of $\ell_{ij}$. Thus, we assume a zero-mean Gaussian prior distribution with precision parameter $\lambda_{ij}$ for $w_{ij}$ as $p(w_{ij})=\cal{N}\mathrm{(w_{ij};0,\lambda^{-1}_{ij})}$, where $\lambda_{ij}$ is learned from the data.
%\begin{equation}
%p(w_{ij})=\mathrm Exp(\lambda_{ij})=\lambda_{ij}\exp(-\lambda_{ij} w_{ij})
%\end{equation}
For diagonal elements of the Laplacian matrix $\Lb$, we have: $\ell_{ii}=\sum_{j\neq i} {w_{ij}}$. Thus, this elements can be obtained from non-diagonal elements of the Laplacian matrix. About the noise vector $\eb$, it is assumed to be zero mean white Gaussian noise with precision $\alpha_e$. So, we have $p(\eb)=\cal{N}\mathrm{(0,\alpha^{-1}_e\Ib_{NK})}$.
%\begin{equation}
%p(\eb)=\cal{N}\mathrm{(0,\alpha^{-1}_e\Ib_{NK})}
%\end{equation}
Also, for the precision parameter $\alpha_e$, a Gamma distribution is assumed. So, we have $p(\alpha_e)=\mathrm{Gamma}(\rho_e,\xi_e)$ where $\rho_e$ and $\xi_e$ are the hyperparameters of $\alpha_e$. Following (\ref{eq: model}), we have

%\begin{equation}
%p(\alpha_e)=\mathrm{Gamma}(\rho_e,\xi_e)
%\end{equation}
\begin{equation}
p(\yb|\xb)=\cal{N}\mathrm{(\Psi\xb,\alpha^{-1}_e\Ib_{NK})}
\end{equation}

\subsection{The proposed VB algorithm}
In this part, the VB inference algorithm is exploited to derive the posterior of the unknown variables and reconstruct the graph signal. Let $\thetab=\{\xb,\Lb,\alpha_e\}$ denote the set of all the unknown variables. Then, in VB, instead of calculating the posterior of the parameter vector $p(\thetab|\yb)$, it is estimated by $q(\thetab)$ in an iterative manner, which minimizes the Kulback-Liebler (KL) divergence between the posterior and the estimated posterior \cite{Beal03}. In general, this estimate is \cite{Tzikas08}:

%\begin{displaymath}
%\ln(q(\theta_i))\propto <\ln(p(\thetab)\mid\yb)>_{\thetab\setminus\theta_i}\propto <\ln(p(\xb,\Lb,\alpha_e\mid\yb)>_{\thetab\setminus\theta_i}
%\end{displaymath}
%\begin{equation}
%<\ln(\frac {p(\yb,\xb,\Lb,\alpha_e)}{p(\yb)}>_{\thetab\setminus\theta_i}
%\end{equation}

\begin{equation}
\ln(q(\theta_i))= \left<\ln(p(\thetab),\yb)\right
>_{\thetab\setminus\theta_i}+\mathrm{const.}
\end{equation}

where $\left<\right>$ stands for statistical average with respect to its indices, and $\thetab\setminus\theta_i$ stands for all variables except variable $\theta_i$. In VB, we assume that the posterior distribution factorizes into independent factors for $\xb$, $\Lb$,and $\alpha_e$, as
\begin{equation}
q(\theta)= q(\xb)q(\alpha_e)\prod_{i,j>i} q(w_{ij})
\end{equation}
with the given restrictions that $\ell_{ij}=-w_{ij}$ for $i\neq j$, $\ell_{ii}=\sum_{j\neq i} {w_{ij}}$, and $\ell_{ij}=\ell_{ji}$. Following the statistical model presented in (\ref{sec: stat}) and the VB approach, the posterior of the graph signal $\xb$ is calculated as
%\begin{displaymath}
%\ln q(\xb)\propto <\ln p(\yb\mid \xb ,\alpha_e)+\ln p(\xb)>_{\Lb,\alpha_e}\propto
%\end{displaymath}
%\begin{displaymath}
%<-\frac{\alpha_e}{2}(\yb-\Psi\xb)^{T}(\yb-\Psi\xb)-\frac{1}{2}\xb^T\Bb\xb>_{\Lb,\alpha_e}\propto
%\end{displaymath}
%\begin{equation}
%\alpha_e\xb^T\Psi^T\yb-\frac{1}{2}\xb^T(\Bb+\alpha_e\Psi^T\Psi)\xb
%\end{equation}

\begin{displaymath}
\ln q(\xb)= \left<\ln p(\yb, \xb, \Lb, \alpha_e)\right>_{\Lb,\alpha_e}+\mathrm{const.}
\end{displaymath}
\begin{displaymath}
=\left<-\frac{\alpha_e}{2}(\yb-\Psi\xb)^{T}(\yb-\Psi\xb)-\frac{1}{2}\xb^T\Bb\xb\right>_{\Lb,\alpha_e}+\mathrm{const.}
\end{displaymath}
\begin{equation}
=\left<\alpha_e\right>\xb^T\Psi^T\yb-\frac{1}{2}\xb^T(\left<\Bb\right>+\left<\alpha_e\right>\Psi^T\Psi)\xb+\mathrm{const.}
\end{equation}

In is clear that this posterior belongs to the Gaussian distribution family $\cal{N}\mathrm{(\mu,\Sigma)}$ with the following mean and covariance:
\begin{displaymath}
\mu=\Sigma\left<\alpha_e\right>\Psi^T\yb
\end{displaymath}

\begin{equation}
\label{eq_signal}
\Sigma=(\left<\Bb\right>+\left<\alpha_e\right>\Psi^T\Psi)^{-1}
\end{equation}
Moreover, since the Gamma distribution is a conjugate prior of the Gaussian distribution, the posterior of the precision of noise has also Gamma distribution as
\begin{displaymath}
q(\alpha_e)=\mathrm{Gamma}(\rho^{'}_e,\xi^{'}_e)
\end{displaymath}
\begin{displaymath}
\rho^{'}_e=\rho_e+\frac{NK}{2}
\end{displaymath}
\begin{displaymath}
\xi^{'}_e=\xi_e+\frac{1}{2}\left<(\yb-\Psi\xb)^T(\yb-\Psi\xb)\right>
\end{displaymath}
\begin{equation}
=\xi_e+\frac{1}{2}(\yb-\Psi\mu)^T(\yb-\Psi\mu)+\frac{1}{2}\mathrm{Tr}(\Psi\Sigma\Psi^T)
\end{equation}
Therefore, at each VB iteration, the updated gamma distribution $q(\alpha_e)$ is used to calculate its expectation $\left<\alpha_e\right>=\frac{\rho^{'}_e}{\xi^{'}_e}$, which is then used in the update of the other parameters.
%\begin{equation}
%\label{eq_noise}
%\alpha_e=\frac{\rho^{'}_e}{\xi^{'}_e} .
%\end{equation}
For the elements of Laplacian matrix $\ell_{ij}$, we can derive the posterior distribution of the elements of the weighted adjacency matrix $w_{ij}$, and exploit the relationship between $\Lb$ and $\Wb$. So we have

%\begin{displaymath}
%\ln q(w_{ij})\propto <\ln p(\xb\mid\Wb)+\ln p(w_{ij})>_{\xb,\alpha_e,\Wb\setminus w_{ij}}\propto
%\end{displaymath}
%\begin{displaymath}
% <\ln p(\xb\mid\L)>_{\xb,\alpha_e,\Lb\setminus \ell_{ij}}+<\ln p(w_{ij})> \propto
%\end{displaymath}
%\begin{displaymath}
%<\ln(\mathrm{det}(\Bb))-\frac{1}{2}\xb^T\Bb\xb>_{\xb,\alpha_e,\Lb\setminus \ell_{ij}}+<\ln(p(w_{ij}))>\propto
%\end{displaymath}
%\begin{displaymath}
%<\ln(\mathrm{det}(\Lb+\epsilon \Ib)^\frac{K}{2})-\frac{1}{2}\sum_{k}x_{k,i}x_{k,j}\ell_{ij}>_{\xb,\alpha_e,\Lb\setminus \textcolor{blue}{\ell_{ij}}}+
%\end{displaymath}
%\begin{equation}
%<\ln(p(\ell_{ij}))>
%\end{equation}
%

\begin{displaymath}
\ln q(w_{ij})= \left<\ln p(\xb\mid\Wb)+\ln p(\Wb)\right>_{\xb,\alpha_e,\Wb\setminus w_{ij}}+\mathrm{const.}
\end{displaymath}
\begin{displaymath}
 =\left<\ln p(\xb\mid\Lb)+\ln p(w_{ij})\right>_{\xb,\Wb\setminus w_{ij}} +\mathrm{const.}
\end{displaymath}
\begin{displaymath}
=\left<\ln((\mathrm{det}(\Bb))^{\frac{1}{2}})-\frac{1}{2}\xb^T\Bb\xb+\ln(p(w_{ij}))\right>_{\xb,\Wb\setminus w_{ij}}+\mathrm{const.}
\end{displaymath}
\begin{displaymath}
=\left<\ln(\mathrm{det}(\Lb+\epsilon \Ib))^\frac{K}{2})-\sum_{k}x_{k,i}x_{k,j}\ell_{ij}+\ln(p(w_{ij}))\right>_{\xb,\Wb\setminus {w_{ij}}}
\end{displaymath}
\begin{displaymath}
+\mathrm{const.}=\frac{K}{2}\left<\ln(\mathrm{det}(\Lb+\epsilon \Ib))\right>_{\Wb\setminus {w_{ij}}}+\sum_{k}\left<x_{k,i}x_{k,j}\right>_{\xb}w_{ij}
\end{displaymath}
\begin{equation}
\label{eq_w}
-\frac{\lambda_{ij}}{2}w^2_{ij}+\mathrm{const.}
\end{equation}

where we have
\begin{equation}
\left<x_{k,i}x_{k,j}\right>_{q(\xb)}=\mu_{k,i}\mu_{k,j}+[\Sigma]_{k,ij}
\end{equation}
To calculate the first term of the penultimate line in (\ref{eq_w}), we parameterize the $(i,j)$'th and $(j,i)$'th elements (due to the symmetrical property of the Laplacian matrix $\Lb$ ($\ell_{ij}=\ell_{ji}$)) with $\ell_{ij}$ and then, calculate the determinant, which results in a second order term with respect to non-diagonal elements as follows
\begin{equation}
\mathrm{det}(\Lb+\epsilon \Ib)=c_{ij}\ell^2_{ij}+\ell_{ij}d_{ij}+g_{ij}
\end{equation}
where the constants $c_{ij}$ and $d_{ij}$ are the factors of $\ell^2_{ij}$ and $\ell_{ij}$ in the determinant term, respectively, and $g_{ij}$ is the constant term in the determinant which is independent of $\ell_{ij}$. Therefore, replacing $\ell_{ij}$ with $-w_{ij}$, we have
\begin{equation}
\mathrm{det}(\Lb+\epsilon \Ib)=c_{ij}w^2_{ij}-w_{ij}d_{ij}+g_{ij}
\end{equation}
which can be rewritten as
\begin{displaymath}
\mathrm{det}(\Lb+\epsilon \Ib)=c_{ij}(w_{ij}-\frac{d_{ij}}{2c_{ij}})^2+c_{ij}(\frac{g_{ij}}{c_{ij}}-\frac{d^2_{ij}}{4c^2_{ij}})=
\end{displaymath}
\begin{equation}
c_{ij}(w_{ij}-u_{ij})^2+c_{ij}z_{ij}
\end{equation}
where $u_{ij}=\frac{d_{ij}}{2c_{ij}}$ and $z_{ij}=\frac{g_{ij}}{c_{ij}}-\frac{d^2_{ij}}{4c^2_{ij}}$. So, we have:
%\begin{displaymath}
%\ln q(w_{ij})\propto \frac{K}{2}\ln[c_{ij}(w_{ij}-u_{ij})^2+c_{ij}z_{ij}]
%\end{displaymath}
%\begin{equation}
%+\sum_{k}x_{k,i}x_{k,j}w_{ij}-\lambda_{ij}\ell_{ij}
%\end{equation}
\begin{displaymath}
\ln q(w_{ij})= \frac{K}{2}\ln[c_{ij}(w_{ij}-u_{ij})^2+c_{ij}z_{ij}]
\end{displaymath}
\begin{equation}
+\sum_{k}(\mu_{k,i}\mu_{k,j}+[\Sigma]_{k,ij})w_{ij}-\frac{\lambda_{ij}}{2}w^2_{ij}+\mathrm{const.}
\end{equation}

By substituting $\lambda_{ij}=\frac{\sum_{k}(\mu_{k,i}\mu_{k,j}+[\Sigma]_{k,ij})}{u_{ij}}$, it can be shown that
\begin{displaymath}
\ln q(w_{ij})= \frac{K}{2}\ln[c_{ij}(w_{ij}-u_{ij})^2+c_{ij}z_{ij}]
\end{displaymath}
\begin{equation}
-\frac{\lambda_{ij}}{2}(w_{ij}-u_{ij})^2+\mathrm{const.}
\end{equation}
 where the above posterior distribution is the generalization of the compound confluent hypergeometric (CCH) distribution, called Generalized CCH (GCCH) distribution, as given by
%\begin{displaymath}
%q(w_{ij})=
%\end{displaymath}
\begin{equation}
q(w_{ij})=\mathrm{GCCH}(w_{ij};2,0.5,\frac{K}{2}+1,0,\frac{\lambda_{ij}}{2},(-z_{ij})^{-0.5},0,u_{ij})
\end{equation}
The GCCH distribution and its explanation is defined in the appendix. According to (\ref{eq_mean_GCCH}), and the relation between $\ell_{ij}$ and $w_{ij}$, the updated GCCH distribution is used to calculate its expectation as
\begin{equation}
\label{eq_GCCH}
\left<\ell_{ij}\right>=-(u_{ij}+\frac{0.5}{\frac{K}{2}+1.5}\frac{\mathrm{H}(1.5,\frac{K}{2}+1,0,\frac{\lambda_{ij}}{2},(-z_{ij})^{-0.5},0)}{\mathrm{H}(0.5,\frac{K}{2}+1,0,\frac{\lambda_{ij}}{2},(-z_{ij})^{-0.5},0)})
\end{equation}
where the function $\mathrm{H}$ is given by (\ref{eq_H}). The GCCH distribution has many pdf's as its special case. In particular, in (\ref{eq_GCCH}), if $c_{ij}\neq0$ and $z_{ij}\neq0$, then we have the GCCH distribution for the corresponding $\ell_{ij}$ as presented in (\ref{eq_GCCH}); if $c_{ij}\ne 0$ and $z_{ij}=0$, or if $c_{ij}=0$ and $d_{ij}\neq0$, then the GCCH distribution simplifies to the three-parameter Gamma distribution, if $c_{ij}=d_{ij}=0$, and $g_{ij}\neq0$, then we have the exponential distribution; and, finally, the case of $c_{ij}=d_{ij}=g_{ij}=0$, results a zero value for the corresponding $\ell_{ij}$. Finally, diagonal elements of the Laplacian matrix can be obtained as $\ell_{ii}=-\sum_{j\neq i}\ell_{ij}$.
%\begin{equation}
%\label{eq_diag}
%\ell_{ii}=-\sum_{j\neq i}\ell_{ij}
%\end{equation}
The overall VB algorithm for graph signal recovery is summarized in Algorithm \ref{Algorithm_1}. As it is evident from this pseudo-code, the VB procedure iteratively updates the parameters and hidden variables until a fixed point is reached or the number of iterations exceeds a predefined number. Moreover, to update the parameters through the VB procedure, the computational complexity of the proposed method is $O(K^3N^3)$ per iteration.

\begin{algorithm}[tb]
\caption{Proposed VB-based graph signal inference algorithm}
\textbf{Input}   \textbf{Observations} $\yb$; \textbf{Sampling matrix} $\Psi$. \newline
\textbf{Initialize} $\hat{\xb}=\Psi^{T}\yb, \lambda_{ij}, \rho_e, \xi_e$.
\label{Algorithm_1}
\begin{algorithmic}
\REPEAT
\begin{itemize}
\item Update the elements of Laplacian matrix $\ell_{ij}$ $(1\le i,j\le N)$ using \eqref{eq_GCCH} for $i\neq j$ and $\ell_{ii}=-\sum_{j\neq i}\ell_{ij}$
\item Update the expectation of posterior of the noise precision $q(\alpha_e)$ via $\left<\alpha_e\right>=\frac{\rho^{'}_e}{\xi^{'}_e}$ and then set $\sigma^2_e=\alpha^{-1}_e$
\item Update the mean and covariance of signal using \eqref{eq_signal} and update the expectation of posterior of signal $q(\xb)$ by its mean
\end{itemize}
\UNTIL {A stopping criterion is reached}
\end{algorithmic}
\end{algorithm}
\vspace{-2ex}

\section{Simulation Results}
\label{sec:Simulation}
In this section, the performance of the proposed VB-based graph signal inference algorithm is evaluated in comparison with the state-of-the-are techniques for real and synthetic data. The performance of the proposed method is evaluated by comparing the normalized mean square error (NMSE)
\begin{displaymath}
\mathrm{NMSE}=\frac{1}{K}\sum_{k=1}^K \frac{||\hat{\xb}_k-\xb_k||^{2}_2}{||\xb_k||^{2}_2}
\end{displaymath}

\begin{figure}[!t]
\vspace{-7ex}
\centering
\hspace*{-1em} \subfloat[NMSE for graph signal]{\includegraphics[width=1.6in]{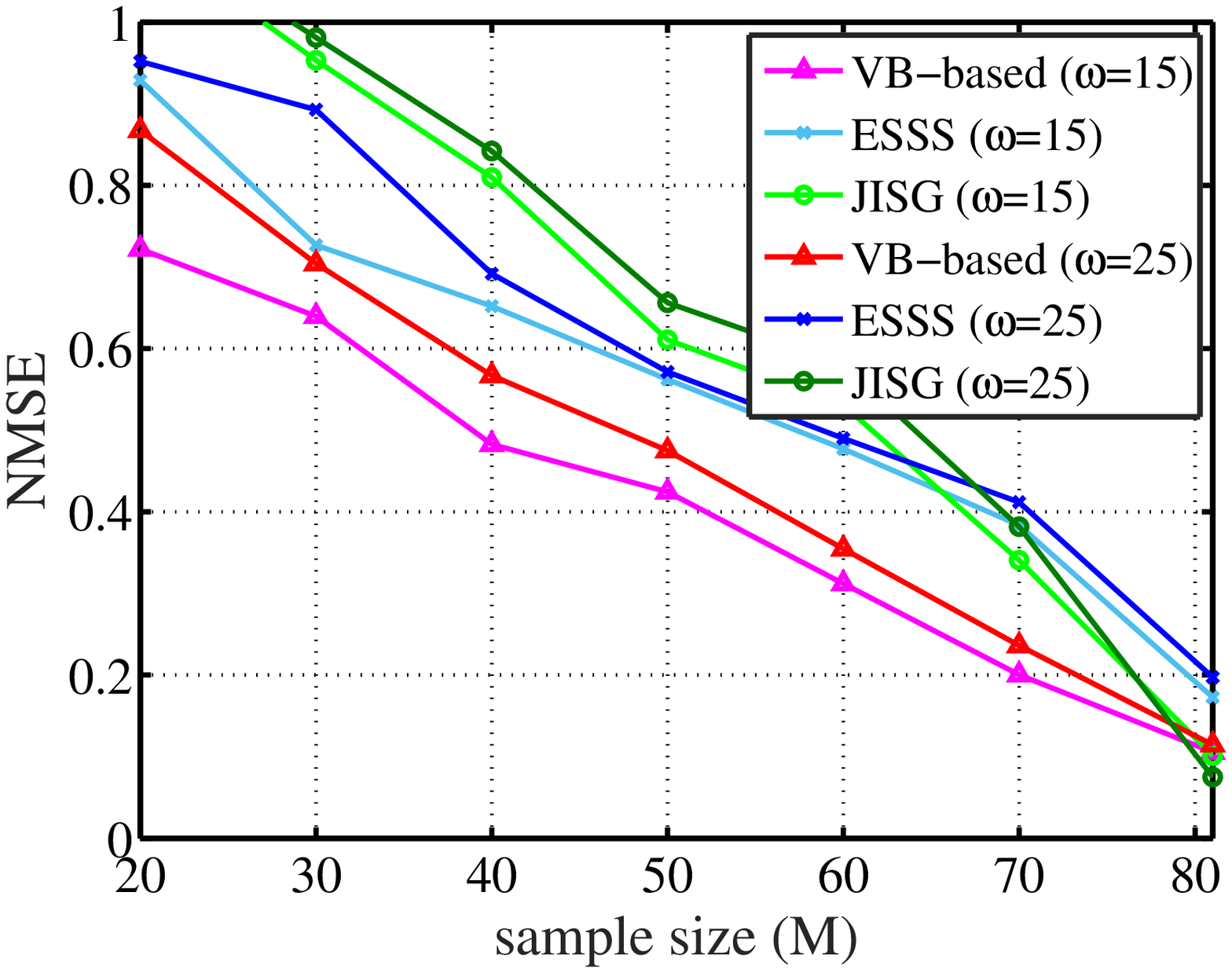}\label{signal_1} } \hspace*{-1.4em} \subfloat[NMSE for Laplacian]{\includegraphics[width=1.6in]{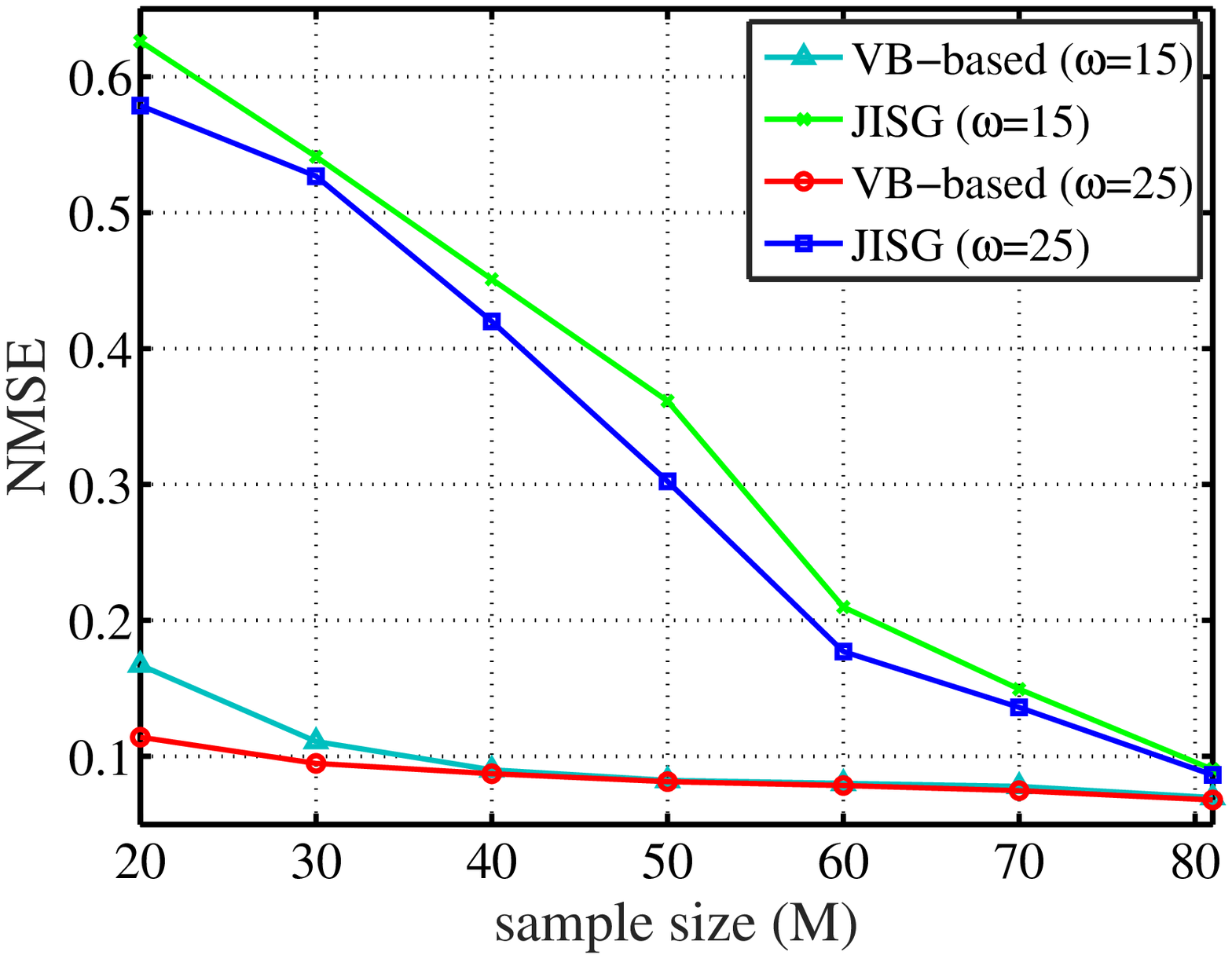}\label{laplacian_1} }
\caption{The NMSE of the GSR for synthetic data ($N=81$ and $K=20$).}
\end{figure}

The reconstruction quality performance of our proposed algorithm is compared with the two approaches: joint inference of the network topology and signals over graphs (JISG) \cite{Ioan19}; and efficient sampling set selection for bandlimited graph signals (ESSS) \cite{Anis16}. In all experiments, the value of parameters selected as $\lambda_{ij}=10^{-2}$, $\rho_e=10^{-6}$, $\xi_e=10^{-6}$, and $\epsilon=10^{-2}$. Also, without loss of generality, we have assumed $\Ab_k=\Ab, \forall k$.

\subsection{Experiments on synthetic data}
\label{subsec:synthetic}
First, the performance of the proposed VB-based algorithm is compared to JISG and ESSS on a synthetic graph with the following scenario (similar to the scenario used in \cite{Ioan19}): Consider a network with $N=81$ nodes, which is based on the ``seed matrix''
\begin{displaymath}
\Pb_{0}=
\begin{bmatrix}
0.6 & 0.1 & 0.7 \\
0.3 & 0.1 & 0.5\\
0 & 1 & 0.1
\end{bmatrix}
\end{displaymath}

\begin{figure}[tb]
\begin{center}
\includegraphics[width=5cm]{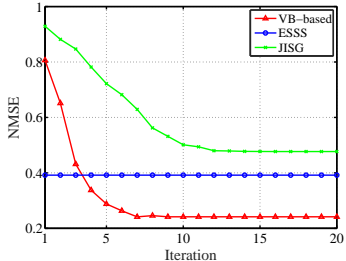}
\end{center}
\caption{Convergence comparison for the synthetic data for $M=70$ and $\omega=15$.}
%\end{center}
\label{convergence}
\end{figure}

and results to an $N\times N$ matrix using the Kronecker product model as $\Pb = \Pb_{0}\otimes \Pb_{0}\otimes \Pb_{0}\otimes \Pb_{0}$. The weighted adjacency matrix $\Wb$ is generated by selecting its elements as $W_{ij}\sim \mathrm{Bernoulli}(P_{ij})$,  $\forall i,j$, and symmetrizing by adding its transpose. Also, the Laplacian matrix is computed using $\Lb=\mathrm{diag}(\Wb.\onevec_N)-\Wb$. To generate each graph signal, we use the graph-bandlimited model $\yb_k=\sum_{i=1}^{\omega} \gamma^{(i)}_k \rb^{(i)}$, $(k=1,...,K)$, where $K=20$, $\gamma^{(i)}_k\sim\cal{N}\mathrm{(0,1)}$, $\omega$ is the bandwidth parameter, and $\{\rb^{(i)}\}^{\omega}_{i=1}$ denotes the eigenvectors corresponding to the $\omega$ smallest eigenvalues of the Laplacian matrix. Finally, we add the Gaussian noise $\eb_k$ to each graph signals, whose elements are drawn from $\cal{N}\mathrm{(0,\sigma_n^2)}$. The variance of noise is selected such that yields $\mathrm{SNR}\sim 6\mathrm{dB}$. Note that, to make a fair comparison, only the non-dynamic part of the JISG algorithm is included. Also, the parameters of JISG are tuned as presented in \cite{Ioan19}. Fig. \ref{signal_1} shows the NMSE vs. the number of selected nodes $M$. From these results, we observe that the VB-based graph signal recovery algorithm consistently performs better than the other algorithms, both for the $\omega =15$ and $25$. Moreover, it can be observed that when $M=N$, all the algorithms perform well. However, when the number of selected nodes becomes lower, the proposed VB-based algorithm performs particularly better than the other methods. The NMSE results for the reconstruction of the Laplacian matrix is presented in fig. \ref{laplacian_1}. As it is evident from this figure, the proposed VB-based algorithm has better graph topology estimation,  both for the $\omega =15$ and $25$. Moreover, the convergence rates of the competing methods is illustrated in fig. \ref{convergence} for $M=70$, $\omega=15$ and $\mathrm{SNR}\sim 6\mathrm{dB}$. We observe that the proposed algorithms becomes stable after around 3 iterations, which proves the fast convergence of our algorithm.

\begin{figure}[!t]
\vspace{-12.1ex}
\centering
\hspace*{-1em} {\includegraphics[width=1.6in]{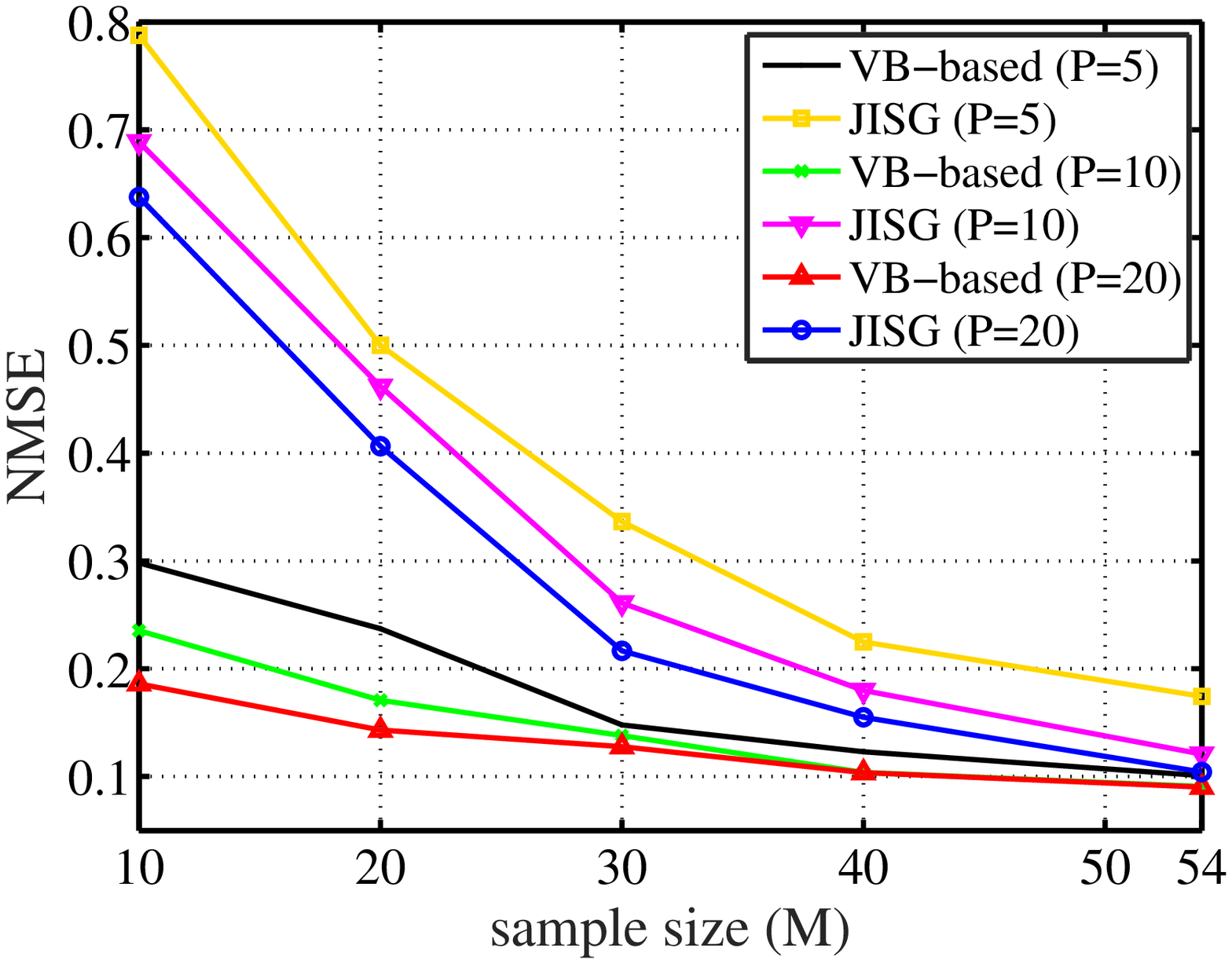}\label{temperature_1} } \hspace*{-1.4em} {\includegraphics[width=1.6in]{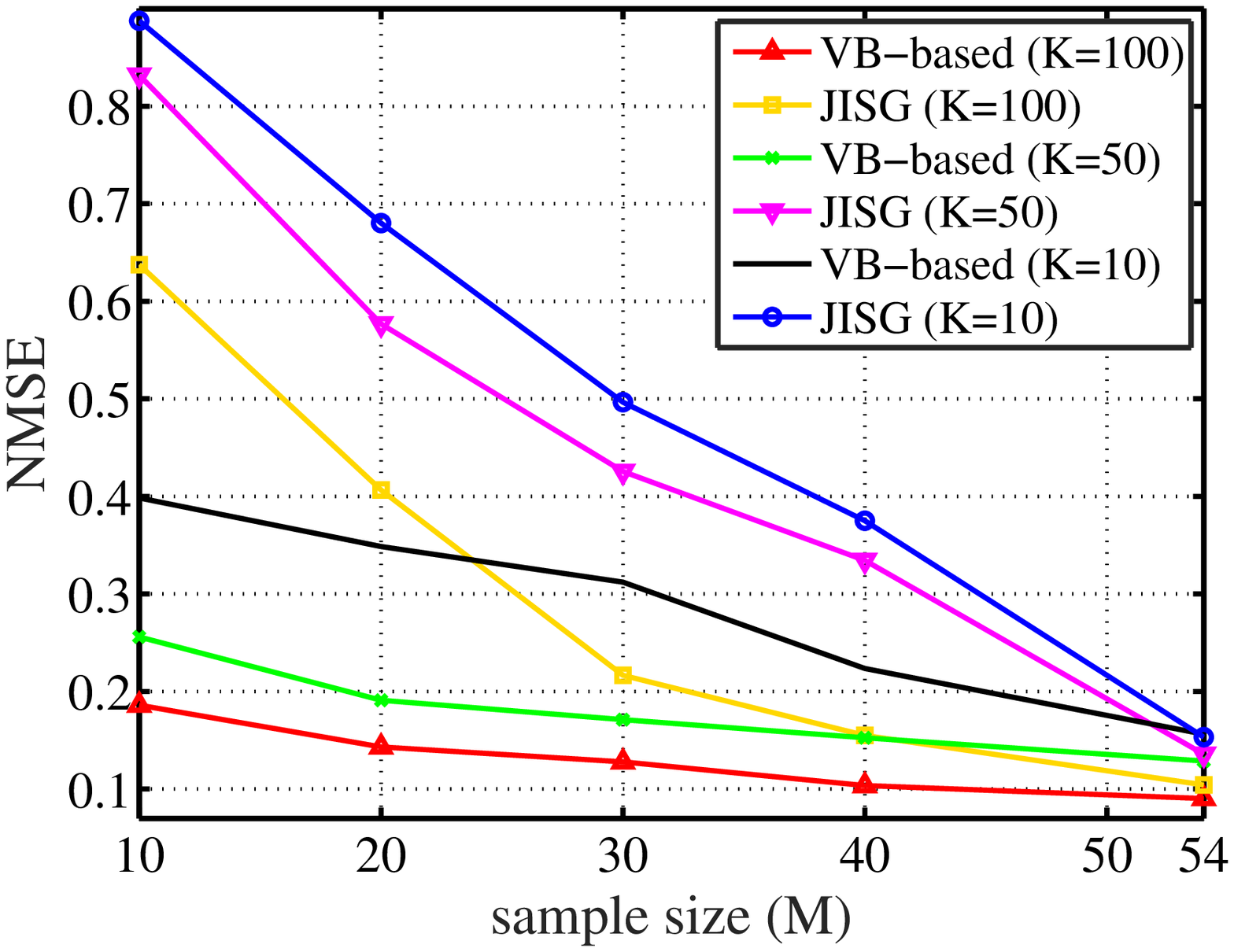}\label{temperature_2}  }
\caption{NMSE for the temperature data.}
%\end{center}
\label{temperature1}
\end{figure}

\subsection{Experiments on temperature data}
\label{subsec:temperature}
In the second scenario, we test the efficiency of our proposed algorithm for 1-D temperature signals downloaded from the Intel  Berkeley Research lab \cite{Intel04}, which contains temperature measurements collected from $N=54$ sensors.
In this experiment, we make a comparison between the proposed VB-based algorithm and non-dynamic part of the JISG method. In this experiment, a P-connected Harary graph \cite{Baudon12}  with $P=5$, $10$ and $20$ is applied to form the neighborhood relationship, where each node is only allowed to communicate information with its P neighbors. For comparing the efficiency of our proposed algorithm in estimating the graph signal, Fig. \ref{temperature_1} plots the NMSE results versus the number of selected nodes $M$, for $\mathrm{SNR}\sim 7\mathrm{dB}$, and for $K=100$. As can be seen, the reconstruction error obtained using VB-based algorithm is smaller than that of the JISG algorithm. Moreover, increasing the number of connections between nodes results in an improvement of reconstruction error.  Fig. \ref{temperature_2} plots the NMSE results for three different sizes of the observation signals at each node (K = 10, 50, 100). It is evident that increasing the number of observations results in an improvement of reconstruction error. These improvements in signal recovery are due to the utilization of a Bayesian method, which exactly models all the statistical behaviors of the unknowns.

\section{Conclusion}
\label{sec:conclusion}
In this paper, we explored the problem of graph signal recovery when the Laplacian matrix of the graph is unknown and its elements are statistically modeled by Gaussian distribution. We modelled the graph signal with a statistical GMRF model. Then, the problem of GSR is solved via a VB approach in a Bayesian manner. In the derivations of the VB-based algorithm, all of the posteriors are calculated in a closed form. Moreover, the posteriors of the Laplacian matrix elements are proved to have a new distribution which is the generalization of the CCH distribution. Finally, the simulation experiments with synthetic data and real data shows the advantage of the proposed Bayesian approach over three state-of-the-art algorithms.

\begin{appendices}
\section{Generalized CCH distribution}
The ``compound confluent hypergeometric'' (CCH) distribution, generalizes and unifies three generalizations of the beta distribution: the generalized beta (GB) distribution \cite{Gordy98}, the Gauss hypergeometric (GH) distribution, and the confluent hypergeometric (CH) distribution. The density function of the CCH distribution is given by \cite{Gordy98}
\begin{displaymath}
\mathrm{CCH}(x;p,q,r,s,v,\theta)=x^{p-1}(1-vx)^{q-1}\times
\end{displaymath}
\begin{equation}
\frac{(\theta+(1-\theta)vx)^{-r}\mathrm{exp}(-sx)}{\mathrm{B}(p,q)\mathrm{H}(p,q,r,s,v,\theta)}
\end{equation}
where $\mathrm{B}$ is the beta function, and $\mathrm{H}$ is given by
\begin{equation}
\label{eq_H}
\mathrm{H}(p,q,r,s,v,\theta)=v^{-p}\mathrm{exp}(\frac{-s}{v})\Phi_1(q,r,p+q,\frac{s}{v},1-\theta)
\end{equation}
where $\Phi_1$ is the confluent hypergeometric function of two variables defined by \cite{Gordy98}
\begin{equation}
\Phi_1(\alpha,\beta,\gamma,x,y)=\sum_{m=0}^{\infty}\sum_{n=0}^{\infty}{\frac{(\alpha)_{m+n}(\beta)_n}{(\gamma)_{m+n}m!n!}x^{m}y^{n}}
\end{equation}
and where $(a)_k$ is Pochhammer's notation
\begin{equation}
(a)_k=
\begin{cases}
1 & \text{if } k=0,\\
a(a+1)(a+2)\cdots(a+k-1) & \text{if } k\neq0.
\end{cases}
\end{equation}
This paper, introduces a generalized version of the CCH distribution, named generalized CCH (GCCH) distribution, which is defined by the pdf
\begin{displaymath}
\mathrm{GCCH}(x;\alpha,p,q,r,s,v,\theta,u)=(x-u)^{\alpha p-1}\times
\end{displaymath}
\begin{displaymath}
(1-v(x-u)^\alpha)^{q-1}(\theta+(1-\theta)v(x-u)^\alpha)^{-r}\times
\end{displaymath}
\begin{equation}
\frac{\mathrm{exp}(-s(x-u)^\alpha)}{\mathrm{B}(p,q)\mathrm{H}(p,q,r,s,v,\theta)}
\end{equation}
Moreover, the mean of the GCCH distribution is given by
\begin{equation}
\label{eq_mean_GCCH}
\mathrm{E}(x)=u+\frac{p}{p+q}\frac{\mathrm{H}(p+\frac{1}{\alpha},q,r,s,v,\theta)}{\mathrm{H}(p,q,r,s,v,\theta)}
\end{equation}

\end{appendices}

\end{document}